\documentclass{amsart}

\usepackage[table,xcdraw]{xcolor}
\theoremstyle{definition}

\usepackage{kotex}

\theoremstyle{remark}

\numberwithin{equation}{section}
\usepackage{url}
\usepackage{amssymb}
\usepackage{graphicx} 
\usepackage{subcaption}
\usepackage{hyperref}
\usepackage{listings}
\usepackage{graphicx}
\usepackage{caption}
\usepackage{subcaption}

\usepackage{xcolor} 
\graphicspath{ {/Users/ekim/Desktop/Stuff} }
\begin{document}
	
	\title{Predicting Participation in Cancer Screening Programs with Machine Learning}
	
	\author{Donghyun (Ethan) Kim}
	\address{Gyeonggi Suwon International School, 451 YeongTong-Ro, YeongTong-Gu, Suwon-Si, Gyeonggi-Do, Republic of Korea}
	
	\email{ethank11k@gmail.com}

	

	\dedicatory{{\normalfont Gyeonggi Suwon International School, 451 YeongTong-Ro, YeongTong-Gu, Suwon-Si, Gyeonggi-Do, Republic of Korea}}

	\begin{abstract}
		In this paper, we present machine learning models based on random forest classifiers, support vector machines, gradient boosted decision trees, and artificial neural networks to predict participation in cancer screening programs in South Korea. The top performing model was based on gradient boosted decision trees and achieved an area under the receiver operating characteristic curve (AUC-ROC) of 0.8706 and average precision of 0.8776. The results of this study are encouraging and suggest that with further research, these models can be directly applied to Korea's healthcare system, thus increasing participation in Korea’s National Cancer Screening Program.  

	\end{abstract}
	
	\maketitle
	
	
	
	
	\section{Introduction} 
	
	In South Korea, cancer is the leading cause of death. In fact, in 2018, cancer was responsible for 26.5\% of all deaths in the country (\cite{K}). As the early detection and diagnosis of cancers increase a patient's change of survival considerably, the Korean government established a National Cancer Screening Program, covering 6 major cancers, including gastric cancer, colorectal cancer, breast cancer, cervical cancer, liver cancer, and lung cancer. To increase participation in this program, the government offers free screening tests for National Health Insurance (NHI) beneficiaries in the lower 50\% income bracket; for those in the upper 50\% income bracket, 90\% of associated costs are covered by the NHI (\cite{B}).

    However, even with such policies, the participation rate in cancer screening programs was 55.6\% in 2019 and showed minimal improvement from 50.1\% in 2015 (\cite{A}).
    
    In this paper, we aim to expand upon previous research conducted on the factors associated with participation in cancer screening programs by developing machine learning models to predict participation in cancer screening programs.
    
    First, section 2 will discuss related studies and associated factors. Next, section 3 will present the method with utilized data, selected variables, data pre-processing steps, and chosen algorithms. Finally, sections 4 will present experimental results and sections 5 and 6 will discuss the results and any implications of the findings.

	\section{Related Studies}

    A number of studies have been conducted on the factors associated with cancer screening participation. With various statistical tools and models, these studies found that factors including education level, income level, and smoking habits are significantly correlated with participation in cancer screening (\cite{C},\cite{D},\cite{E}).
    
    However, very few studies have been conducted on the application of machine learning to predict cancer screening participation. One particular study utilized a hybrid neural network to predict breast screening attendance (breast cancer) in the United Kingdom and achieved 80\% accuracy for the algorithm (\cite{F}). Another study utilized machine learning (support vector machines, random forests, etc.) to predict hospital attendance and achieved 0.852 for the area under receiver operating characteristic curve (\cite{G}).

	\section{Method}

	\subsection{Data}
	
	Data was obtained from the Seventh Korea National Health and Nutrition Examination Survey, including data from 2016 \- 2018. 24269 individuals participated in the survey (10611 households) and participants were chosen through stratified cluster sampling. More specifically,  primary sampling units were selected based on results from the annual Population and Housing Census.
	
	Available data for the 3 years ranges from demographic information to dietary habits and includes a total of 852 variables (\cite{O}).
	
	\subsection{Features/Variables}
	
	Variables were selected based on findings from past studies and their relevance were verified with the Chi-Squared test (\cite{C},\cite{D},\cite{E}). The 55 (89 after pre-processing --- one-hot encoding) chosen variables are listed below.
    
\begin{table}[h]
	    \hspace*{-1cm}
	    \resizebox{13.4cm}{!} {
	    
		\begin{tabular}{|l|l|l|} 
			\hline
			Variable Description & Variable Type \\ [0.5ex] 
			\hline\hline
			 Income Level & Ordinal \\ 
			 \hline
			 Education Level & Ordinal \\
			 \hline
			 Self Perception of Health & Ordinal \\
			 \hline
			 High Blood Pressure Diagnosis & Binary \\
			 \hline
			 Hyperlipidemia Diagnosis  & Binary\\
			 \hline
			 Stroke Diagnosis  & Binary \\
			 \hline
			 \vtop{\hbox{\strut Myocardial Infarction/Angina}\hbox{\strut Pectoris Diagnosis}} & Binary \\
			 \hline
			 Myocardial Infarction Diagnosis & Binary \\
			 \hline
			 Angina Pectoris Diagnosis & Binary \\
			 \hline
			 Arthritis Diagnosis & Binary \\
			 \hline
			 Osteoarthritis Diagnosis & Binary \\
			 \hline
			 Rheumatoid Arthritis Diagnosis & Binary \\
			 \hline
			 Osteoporosis Diagnosis & Binary \\
			 \hline
			 Tuberculosis Diagnosis & Binary \\
			 \hline
			 Asthma Diagnosis & Binary \\
			 \hline
			 Diabetes Diagnosis & Binary \\
			 \hline
			 Thyroid Gland disease Diagnosis & Binary \\
			 \hline
			 Stomach Cancer Diagnosis & Binary \\
			 \hline
			 Liver Cancer Diagnosis & Binary \\
			 \hline
			 Colon Cancer Diagnosis & Binary \\
			 \hline
			 Breast Cancer Diagnosis & Binary \\
			 \hline
			 Cervical Cancer Diagnosis & Binary \\
			 \hline
			 Lung Cancer Diagnosis & Binary \\
			 \hline
			 Thyroid Cancer Diagnosis & Binary \\
			 \hline
			 Other-1 Cancer Diagnosis & Binary \\
			 \hline
			 Other-2 Cancer Diagnosis & Binary \\
			 \hline
			 Depression Diagnosis & Binary \\
			 \hline
		\end{tabular}

			 \begin{tabular}{|l|l|l|} 
			 Atopic Dermatitis Diagnosis & Binary \\
			 \hline
			 Allergic Rhinitis Diagnosis & Binary \\
			 \hline
			 Sinusitis Diagnosis & Binary \\
			 \hline
			 Otitis Media Diagnosis & Binary \\
			 \hline
			 Cataract Diagnosis & Binary \\
			 \hline
			 Glaucoma Diagnosis & Binary \\
			 \hline
			 Macular Degeneration Diagnosis & Binary \\
			 \hline
			 Renal Failure Diagnosis & Binary \\
			 \hline
			 Hepatitis B Diagnosis & Binary \\
			 \hline
			 Hepatitis C Diagnosis & Binary \\
			 \hline
			 Liver Cirrhosis Diagnosis & Binary \\
			 \hline
			 Influenza Vaccination & Binary \\
			 \hline
			 Regular Health Check-up  & Binary \\
			 \hline
			 Limited Daily/Social Life & Binary \\
			 \hline
			 Employment & Binary \\
			 \hline
			 Self Perception of Stress & Ordinal \\
			 \hline
			 Regular Exercise & Binary \\
			 \hline
			 Nutrition Education Status & Binary \\
			 \hline
			 Private Health Insurance & Binary \\
			 \hline
			 Type of Health Insurance  & Nominal \\
			 \hline
			 Occupation Type & Nominal \\
			 \hline
			 Region & Nominal \\
			 \hline
			 Unmet Healthcare Needs & Nominal \\
			 \hline
			 Causes of Unmet Healthcare Needs & Nominal \\
			 \hline
			 Self Perception of Body Image & Ordinal \\
			 \hline
			 Hospital Admission in Past Year & Binary \\
			 \hline
			 Drinking Level & Ordinal \\
			 \hline
			 Smoking Level & Ordinal \\ [1ex]
			\hline
		\end{tabular}
        }
		\caption{Selected Variables --- Description and Type}
		\label{table:table1}

	\end{table}
    \newpage
	
	\subsection{Data Pre-processing}
	
	First, all rows containing unavailable (marked as "unknown") or missing (null) values were removed from the data-set. Next, all ordinal categorical variables were label encoded and all nominal categorical variables were one-hot encoded. All variables were scaled with Min-Max scaling to a fixed range (0 to 1) as well.
	
	Finally, the dataset was shuffled and split into training and test sets (ratio of 80/20 respectively).

	\subsection{Algorithms}
	
	Algorithms chosen for this task include random forest classifiers, support vector machines, gradient boosted decision trees (XGBoost), and artificial neural networks (with back-propagation).
	
	For all algorithms, grid search and 5-fold cross validation were performed to find optimal hyper-parameters. Each model was then evaluated with the held-out test set. Note that all experiments were conducted via Google Colaboratory.
	
	\section{Results}
    
\begin{figure}[]
        \begin{subfigure}[b]{0.49\textwidth}
            \includegraphics[width=\textwidth]{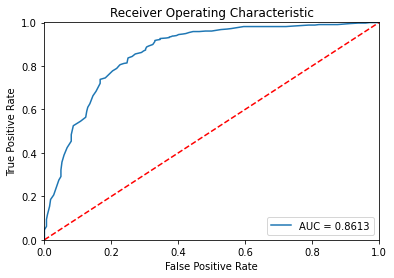}
            \caption{ROC Curve}
            \label{fig:f1}
        \end{subfigure}
        \hfill
        \begin{subfigure}[b]{0.49\textwidth}
            \includegraphics[width=\textwidth]{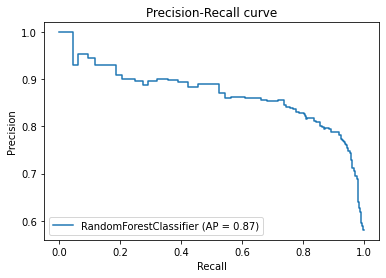}
            \caption{Precision-Recall Curve}
            \label{fig:f2}
        \end{subfigure}
        \caption{Random Forest}
        \label{figure:figure1}
    \end{figure}

\begin{figure}[]
        \begin{subfigure}[b]{0.49\textwidth}
            \includegraphics[width=\textwidth]{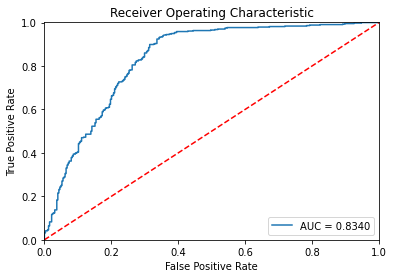}
            \caption{ROC Curve}
            \label{fig:f1}
        \end{subfigure}
        \hfill
        \begin{subfigure}[b]{0.49\textwidth}
            \includegraphics[width=\textwidth]{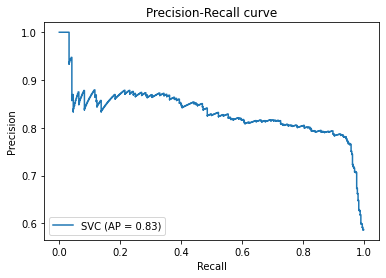}
            \caption{Precision-Recall Curve}
            \label{fig:f2}
        \end{subfigure}
        \caption{Support Vector Machines}
        \label{figure:figure2}
    \end{figure}

\begin{figure}[]
        \begin{subfigure}[b]{0.49\textwidth}
            \includegraphics[width=\textwidth]{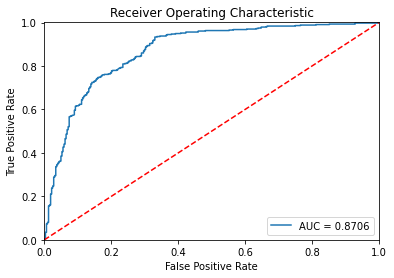}
            \caption{ROC Curve}
            \label{fig:f1}
        \end{subfigure}
        \hfill
        \begin{subfigure}[b]{0.49\textwidth}
            \includegraphics[width=\textwidth]{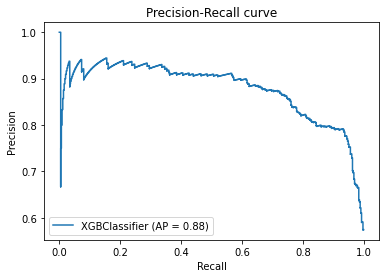}
            \caption{Precision-Recall Curve}
            \label{fig:f2}
        \end{subfigure}
        \caption{XGBoost}
        \label{figure:figure3}
    \end{figure}

\begin{figure}[]
        \begin{subfigure}[b]{0.49\textwidth}
            \includegraphics[width=\textwidth]{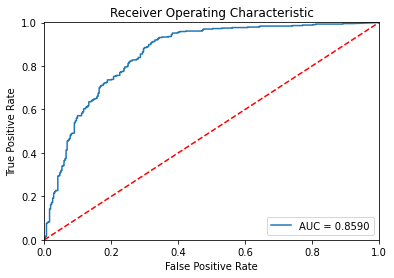}
            \caption{ROC Curve}
            \label{fig:f1}
        \end{subfigure}
        \hfill
        \begin{subfigure}[b]{0.49\textwidth}
            \includegraphics[width=\textwidth]{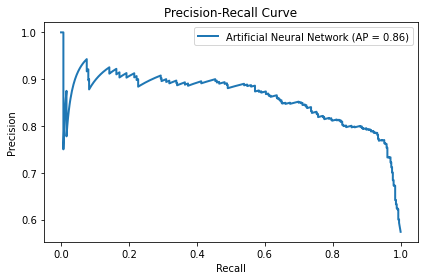}
            \caption{Precision-Recall Curve}
            \label{fig:f2}
        \end{subfigure}
        \caption{Artificial Neural Network}
        \label{figure:figure4}
    \end{figure}

    The performance of the 4 algorithms can be found below in Table \ref{table:table2} and corresponding plots for the Receiver Operating Characteristic curve (ROC) and Precision-Recall curve (PR Curve) can be found in Figures \ref{figure:figure1}, \ref{figure:figure2}, \ref{figure:figure3}, and \ref{figure:figure4}. 
    
    To evaluate each model, 3 accuracy metrics were chosen: Area under the Receiver Operating Characteristic curve (AUC-ROC), Average Precision (Area under the Precision-Recall curve), and Accuracy. Note that AUC-ROC was used as the accuracy metric for hyper-parameter tuning.

	\begin{table}[h]
		\centering
		
		\begin{tabular}{|l|l|l|l|l|} 
			\hline
			Algorithm & AUC-ROC & Average Precision & Accuracy \\ [0.5ex] 
			\hline\hline
			 Random Forest & 0.8613 & 0.8694 & 0.8053 \\
			 \hline
			 Support Vector Machine & 0.8340 & 0.8327 & 0.8118 \\ 
			 \hline
			 XGBoost & 0.8706 & 0.8776 & 0.8171 \\ 
			 \hline
			 Artificial Neural Network & 0.8590 & 0.8605 & 0.8118 \\ [1ex] 
			\hline
		\end{tabular}
		\caption{Accuracy Metrics for Trained Models}
		\label{table:table2}
	\end{table}
    
    \newpage

    \clearpage  

    \section{Discussion}
    
    We see that all 4 models achieved scores between 0.8 and 0.9 for the 3 chosen accuracy metrics. The top performing model was based on gradient boosted decision trees (XGBoost) and achieved an AUC-ROC of 0.8706 and average precision of 0.8776.
    
    One point to note is that the 4 models did not differ significantly in terms of achieved scores. As such, one option to improve prediction accuracy would be to incorporate more features/variables from the original dataset to form more complex models. 

    Furthermore, there was minimal discrepancy between the models' performance during cross-validation and testing, indicating how over-fitting was avoided.
    
    \section{Conclusion}
    
    One limitation of the models developed in this paper is that the Korea National Health and Nutrition Examination Survey contains self-reported data. As such, portions of the data used (especially answers to subjective survey questions) may have been erroneous, which would reduce the accuracy of the model.   
    
    Nevertheless, the models developed in this paper can be directly applied to Korea's healthcare system. Regional public health officials could use these models (or variations of these models --- depending on data availability) to predict individuals who are likely to not participate in cancer screening programs. Officials could then contact these individuals, informing them of screening locations and dates. Both public and private hospitals could make use of these models as well, depending on the availability of data.    
    
    If such a system is implemented, Korea’s National Cancer Screening Program may see a rise in participation.
    
    Further research on this topic with models of greater complexity and additional features may lead to higher prediction accuracy and a rise in overall cancer screening program participation. For instance, varying the total number of features used may lead to more efficient models. The use of complex neural networks with various types of layers may result in models of greater accuracy as well.

	\bibliographystyle{amsplain}

\end{document}